\documentclass[12pt]{article}
\usepackage{graphicx}
\def \s{\sqrt{2}}
\def \bea{\begin{eqnarray}}
\def \eea{\end{eqnarray}}
\def \eeq{\end{equation}}
\def \beq{\begin{equation}}
\def\tl{\tilde\lambda}

\def\lud{\lambda_u^{(d)}}
\def\ltd{\lambda_t^{(d)}}
\def\b{\cal B}

\begin{document}
\rightline{TECHNION-PH-2010-3}
\rightline{EFI 10-6}
%MG changed
\rightline{arXiv:1003.5090v2} 
\bigskip
\centerline{\bf CALCULATING PHASES BETWEEN $B\to K^*\pi$ AMPLITUDES} 
\bigskip 
 
\centerline{Michael Gronau} 
\centerline{\it Physics Department, Technion -- Israel Institute of Technology} 
\centerline{\it 32000 Haifa, Israel} 
\medskip 
 
\centerline{Dan Pirjol} 
\centerline{\it Department of Particle Physics, National Institute for Physics and Engineering} 
\centerline{\it  077125 Bucharest, Romania} 
\medskip

\centerline{Jonathan L. Rosner} 
\centerline{\it Enrico Fermi Institute and Department of Physics,
  University of Chicago} 
\centerline{\it Chicago, IL 60637, U.S.A}.

%JR when adding text please keep it to a maximum of 80 characters per line.

\begin{quote}
A phase $\Delta\Phi$ between amplitudes for $B^0\to K^{*0}\pi^0$ and $B^0\to
K^{*+}\pi^-$ plays a crucial role in a method for constraining
Cabibbo-Kobayashi-Maskawa (CKM) parameters.
We present a general argument for destructive interference between amplitudes
for $B^0\to K^{*+}\pi^-$ and $B^0\to K^{*0}\pi^0$ forming together a smaller
$I(K^*\pi)=3/2$ amplitude. Applying flavor SU(3) and allowing for conservative
theoretical uncertainties, we obtain lower limits on $|\Delta\Phi|$ and its
charge-conjugate. Values of these two phases favored by the Babar collaboration
are in good agreement with our bounds.
\end{quote}
\bigskip

{\bf I. INTRODUCTION}

\medskip

Charmless hadronic $B$ meson decays from $b\to s$ transitions including 
$B\to K\pi$ provide useful information about the weak phase 
$\gamma$~\cite{Gronau:1994bn,Fleischer:1997um,Neubert:1998pt}. 
A method for constraining another angle in the $(\bar\rho, \bar\eta)$ plane, 
formed by the $\bar\rho$ axis and a line going through the apex of the 
%MG added +- 0.03
unitarity triangle intersecting the $\bar\rho$  axis at $\bar\rho = 0.24\pm 0.03$, 
is based on Dalitz analyses of $B^0\to K^+\pi^-\pi^0$ and 
$B^0\to K_S\pi^+\pi^-$~\cite{Ciuchini:2006kv,Gronau:2006qn}. 
The first process enables one to determine a phase $\Delta\Phi$ between 
quasi-two-body decay amplitudes for $B^0\to K^{*0}\pi^0$ and 
$B^0\to K^{*+}\pi^-$, 
\beq
\Delta\Phi \equiv  {\rm Arg}[A(K^{*0}\pi^0)A^*(K^{*+}\pi^-)]~.
\eeq

%MG added in response to referee's comments
While this phase appears as a purely experimental quantity in 
Ref.~\cite{Ciuchini:2006kv,Gronau:2006qn},
the purpose of this work is to obtain bounds on $|\Delta\Phi|$ and its charge
conjugate.  Values of these two phases favored by a recent Babar Dalitz
analysis of $B \to K^\pm \pi^\mp \pi^0$~\cite{Aubert:2008zu,WagnerThesis}
are in agreement with our bounds, once sign conventions for $K^*$ decays are
taken into account.  These results are relevant to extraction of the $I=3/2$
$B \to K^* \pi$ amplitude
%MG added A_{3/2}
$A_{3/2}$, whose phase (along with that of the corresponding
charge-conjugate amplitude) determines the above-mentioned angle in the
$(\bar\rho, \bar\eta)$ plane. 
%MG added sentence
We will shows that as a result of destructive interference found between $A(K^{*0}\pi^0)$
and $A(K^{*+}\pi^-)$ $|A_{3/2}|$ is not well enough known to carry out this program.

This paper will be divided into several short sections. Section II introduces conventions
for defining two quasi-two-body resonant amplitudes, $A(K^{*+}\pi^-)$ and $A(K^{*0}\pi^0)$, 
contributing to $B^0\to K^+\pi^-\pi^0$. In Section III we present a qualitative argument for 
destructive  interference between these two amplitudes forming together an
$I(K^*\pi)=3/2$ amplitude. Crude estimates for $\Delta\Phi$ and its 
charge conjugate $\Delta\overline{\Phi}$ obtained in Section IV assuming flavor SU(3) are improved in Section 
V by including uncertainties from SU(3) breaking and small contributions. Section VI
concludes by comparing our bounds on $\Delta\Phi$ and $\Delta\overline{\Phi}$ with recent 
experimental results obtained by the Babar collaboration.
\bigskip

{\bf II. CONVENTIONS FOR RESONANT AMPLITUDES IN $B^0\to K^+\pi^-\pi^0$}

\begin{figure}
\begin{center}
\includegraphics[height=2in]{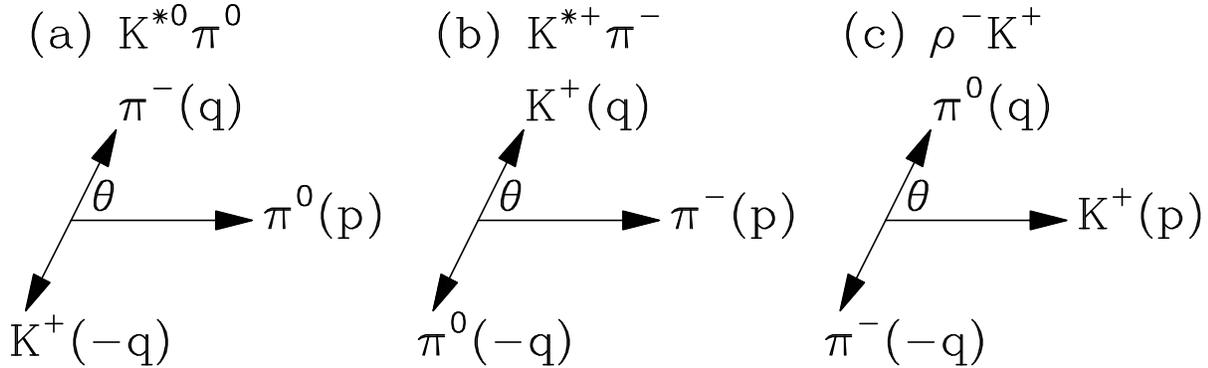}
\end{center}
\caption{Convention of Ref.\ \cite{Aubert:2008zu} for quasi-two-body subsystems
in the three-body decays $B^0 \to K^+ \pi^- \pi^0$.  (a) $B^0 \to K^{*0}
\pi^0$; (b) $B^0 \to K^{*+} \pi^-$; (c) $B^0 \to \rho^- K^+$.
\label{fig:decays}}
\end{figure}

\medskip
The conventions for three-body decays, stated explicitly below Eq.\ (8) of
Ref.\ \cite{Aubert:2008zu}, are illustrated in Fig.\ \ref{fig:decays}.  Each
quasi-two-body subsystem of the three-body decay $B^0 \to K^+ \pi^- \pi^0$, as
viewed in the rest frame of the vector meson, contains pseudoscalar decay
products of the vector meson with momenta $\mathbf{q}$ and $\mathbf{-q}$ and a
bachelor pseudoscalar with momentum $\mathbf{p}$.
% This is Figure 1
Specifically, the phase conventions adopted in Ref.\ \cite{Aubert:2008zu} are
such that (a) for $B^0 \to K^{*0} \pi^0$, the $K^{*0}$ decay particle with
momentum $\mathbf{q}$ is a $\pi^-$, while the bachelor particle with momentum
$\mathbf{p}$ is a $\pi^0$; (b) for $B^0 \to K^{*+} \pi^-$ the $K^{*+}$ decay
product with momentum $\mathbf{q}$ is a $K^+$ while the bachelor particle with
momentum $\mathbf{p}$ is a $\pi^-$; and (c) for $B^0 \to \rho^- K^+$ the
$\rho^-$ decay product with momentum $\mathbf{q}$ is a $\pi^0$ while the
bachelor particle with momentum $\mathbf{p}$ is a $K^+$.  These enter into a
tensor $T = - 2 \mathbf{p \cdot q}$ describing the
matrix element in the Zemach formalism \cite{Zemach}.

One must be careful to use Clebsch-Gordan coefficients appropriate to these
phase conventions when constructing $B^0 \to K^{*0} \pi^0$ and $B^0 \to K^{*+}
\pi^-$ amplitudes from Dalitz-plot fits.  The interchange of the two final-state
particles in $K^* \to K \pi$ causes a sign change as a result of the property
\beq
(j_2 m_2 j_1 m_1|j m) = (-1)^{j-j_1-j_2}(j_1 m_1 j_2 m_2|j m)
\eeq
of the Clebsch-Gordan coefficients \cite{PDG}.  As we choose to use the
same order $(j_1 = 1, j_2 = 1/2)$ in describing both $K^{*0}$ and $K^{*+}$
decays, our relative phases of $A(K^{*+}\pi^-)$ and $A(K^{*0} \pi^0)$ and
their charge-conjugates will be those of Refs.\ \cite{Aubert:2008zu} and
\cite{WagnerThesis} shifted by $180^\circ$.  In our convention a combination
of amplitudes for an $I=3/2$ final $K^* \pi$ state may be written
\beq \label{A3/2}
3 A_{3/2} \equiv A(K^{*+}\pi^-) + \s A(K^{*0} \pi^0),
\eeq
whose magnitude is determined by measuring the magnitudes of the two amplitudes
on the right-hand side and their relative phase.  We will argue, first
qualitatively and then quantitatively, that in the CKM framework these two
amplitudes add destructively in (\ref{A3/2}), implying that the amplitude
$3A_{3/2}$ is smaller in magnitude than either of these two amplitudes.
\bigskip

{\bf III. AN ARGUMENT FOR DESTRUCTIVE INTERFERENCE}

\medskip
Destructive interference in (\ref{A3/2}) follows qualitatively in the CKM
framework from the cancellation of a $\Delta I=0$ penguin amplitude dominating
the two $B\to K^*\pi$ amplitudes on the right-hand side~\cite{Gronau:2010dd}.
The remaining terms on the right-hand side, consisting of electroweak penguin
(EWP) and tree amplitudes, are considerably smaller than the penguin amplitude.
This is demonstrated by decomposing physical amplitudes into graphical
contributions representing distinct flavor topologies~\cite{Gronau:1994rj,%
Dighe:1995gq}, each of which involves an unknown strong phase,
 \bea\label{K*+-}
-A(K^{*+}\pi^-) & = & \lambda_t^{(s)}(P_{tc,P} + \frac{2}{3}P^C_{EW,P})
+ \lambda_u^{(s)}(P_{uc,P} + T_P) ~,\nonumber\\
\label{K*00}
\s A(K^{*0}\pi^0) & = & \lambda_t^{(s)}(P_{tc,P} - P_{EW,V} - \frac{1}{3}P^C_{EW,P})
+  \lambda_u^{(s)}(P_{uc,P} - C_V)~.
\eea
This implies
\beq\label{3A3/2}
3A_{3/2} = -\lambda_t^{(s)}(P_{EW,V} + P^C_{EW,P}) - \lambda_u^{(s)}(T_P + C_V)~.
\eeq
The two CKM factors $\lambda_q^{(q')}\equiv V^*_{qb}V_{qq'}~(q=u, t; q'=d, s)$ have a very 
small ratio $|\lambda_u^{(s)}|/|\lambda_t^{(s)}|\simeq 0.02$~\cite{PDG}. The dominant term 
multiplying $\lambda_t^{(s)}$ in the two $B\to K^*\pi$ amplitudes  is the penguin contribution 
$P_{tc,P}$, while the EWP contributions $P_{EW,V}$ and  $P^C_{EW,P}$ 
are smaller as they are higher order in the electroweak coupling. 
Thus the dominant penguin contributions cancel in $3A^{K^*\pi}_{3/2}$, which consists 
of  two smaller contributions: EWP terms multiplying $\lambda_t^{(s)}$ and a combination of tree 
amplitudes $T_P + C_V$ multiplying a very small CKM factor $\lambda_u^{(s)}$.
\bigskip

{\bf IV. AN APPROXIMATE CALCULATION OF $\Delta\Phi$ and $\Delta\overline{\Phi}$}

\medskip
In order to study quantitatively the interference  between the two $B\to
K^*\pi$ amplitudes in (\ref{A3/2}) we make use of two model-independent
relations:
\begin{itemize}
\item Proportionality relations between tree and EWP operators in the $|\Delta
S| = |\Delta I| = 1$ effective Hamiltonian, in which one neglects EWP operators
${\cal O}_7$ and ${\cal O}_8$ with tiny Wilson coefficients, imply the
following expression for the EWP $I(K^*\pi)=3/2$ amplitude in terms
of tree amplitudes~\cite{Gronau:1998fn,Gronau:2000az},
\beq\label{EWP/T}
P_{EW,V} + P^C_{EW,P} = - \frac{3{\cal K}}{2}(T_V + C_P)~.
\eeq
Here  ${\cal K}$ is a ratio of Wilson coefficients~\cite{Buchalla:1995vs}, 
${\cal K} \equiv (c_9 + c_{10})/(c_1 + c_2) \approx (c_9 - c_{10})/(c_1 - c_2) = -0.0087$. 
\item In the flavor SU(3) limit amplitudes for $B\to \rho\pi$ decays are given in terms of the 
same reduced SU(3) amplitudes (i.e., the same graphical amplitudes) 
contributing to $B\to K^*\pi$, but involve different CKM factors. Thus, neglecting tiny EWP 
amplitudes and annihilation contributions $A_P-A_V$~\cite{Blok:1997yj,Bauer:2004tj}, 
one has 
\bea\label{rho+pi0}
-\s A(\rho^+\pi^0) & = & \lud(T_P + C_V) - \ltd(P_V - P_P)~,\nonumber \\
\label{rho0pi+}
-\s A(\rho^0\pi^+) & = & \lud(T_V + C_P) + \ltd(P_V - P_P)~.
\eea
In the same limit amplitudes for $\Delta S=0$ $B^+\to K^*K$ decays are expressed in terms
of  penguin amplitudes $P_P, P_V$, (again after neglecting small EWP and annihilation contributions),
\bea\label{KK*}
A(\bar K^{*0} K^+) & = &  
 \ltd P_P ~,\nonumber \\
A(K^{*+} \bar K^0) & = &  
\ltd P_V~.
\eea
Here $P_P\equiv P_{tc,P}$ contributes to $B\to K^*\pi$ amplitudes in (\ref{K*00}).
Contributions of  annihilation amplitudes and terms $\lud P_{uc,P}, \lud P_{uc,V}$ 
which have been omitted in (\ref{rho+pi0}) and (\ref{KK*}), respectively,  will be 
included later on.
\end{itemize}

In order to obtain first a rough estimate for $3A_{3/2}$, 
$\Delta\Phi$ and their charge-conjugates we will work 
at this point in the SU(3) symmetry approximation, which is expected to introduce an 
uncertainty of about $20-30\%$ in amplitudes. 
For now we will also neglect  penguin contributions in (\ref{rho+pi0}) 
which can be estimated to be of the same order,
\beq\label{P_P}
\frac{|\ltd P_P|}{| \lud T_P|} \simeq
\sqrt{\frac{{\cal B}(\bar K^{*0}K^+)}{r_\tau{\cal B}(\rho^+\pi^-)}} = 0.20 \pm 0.03~.
\eeq
We have used decay branching ratios and a lifetime ratio, 
$r_\tau\equiv \tau_{B^+}/\tau_{B^0}=1.071\pm 0.009$, from Ref.~\cite{HFAG}. 
Thus we take,
\bea\label{rho+pi0_approx}
-\s A(\rho^+\pi^0) & \simeq & \lud(T_P + C_V)~,\nonumber\\
-\s A(\rho^0\pi^+) &  \simeq & \lud(T_V + C_P)~.
\eea
Flavor  SU(3) symmetry breaking in the amplitudes
$T_P + C_V$ and $T_V + C_P$ and uncertainties caused by neglecting penguin 
amplitudes will be included in the analysis at a later point.

We denote $\tl \equiv \lambda/(1-\lambda^2/2)=0.232$, where $\lambda$ is the Wolfenstein 
parameter~\cite{Wolfenstein:1983yz}, and use the central value for CKM parameters~\cite{PDG},
\beq\label{K}
\frac{3{\cal K}}{2}\frac{\lambda_t^{(s)}}{\lambda_u^{(s)}} = 0.61e^{-i\gamma}~.
\eeq
We checked that uncertainties of $10\%$ in the magnitude of this ratio and a few degrees 
in its strong phase~\cite{Neubert:1998re} have an insignificant effect on the subsequent analysis.
Combining Eqs.~(\ref{3A3/2}), (\ref{EWP/T}), (\ref{rho+pi0_approx}) and (\ref{K}) we obtain 
in this  approximation,
\beq\label{3A3/2approx}
3A_{3/2} \simeq \tl\s\left( A(\rho^+\pi^0) - 0.61e^{-i\gamma}A(\rho^0\pi^+)\right)~.
\eeq
We will now use this approximate expression in order to evaluate the magnitude of 
$3A_{3/2}$ and its CP-conjugate.

%This is Table 1
\begin{table}
\caption{Branching fractions and CP asymmetries for $B\to K^*\pi, \rho\pi$.
For $B\to K^{*+}\pi^-$ we calculate averages of recent Babar 
measurements~\cite{WagnerThesis} and Belle measurements~\cite{Garmash:2006fh},
for $B^0\to K^{*0}\pi^0$ we take values from Ref.~\cite{WagnerThesis} as 
Belle has so far obtained only a loose upper limit on this mode~\cite{Chang:2004um}, while 
for $B\to \rho\pi$ we quote values in~\cite{HFAG}.
\label{tab:k*pirhopi}}
\begin{center}
\begin{tabular}{c c c} \hline \hline
Mode & $\b$~$(10^{-6})$ & $A_{CP}$  \\ \hline\hline
$B^0\to K^{*+}\pi^-$ & $8.2 \pm 1.0$ & $-0.26 \pm 0.08$\\
$B^0\to K^{*0}\pi^0$ & $3.3\pm 0.6$ & $-0.15\pm 0.13$\\ \hline
$B^+ \to \rho^+\pi^0$ & $10.9^{+1.4}_{-1.5}$ & $0.02\pm 0.11$ \\
$B^+ \to \rho^0 \pi^+$ & $8.3^{+1.2}_{-1.3}$  & $0.18^{+0.09}_{-0.17}$\\
\hline \hline
\end{tabular}
\end{center}
\end{table}

CP-averaged branching ratios and CP asymmetries for relevant $B\to K^*\pi$ and $B\to\rho\pi$
decays are given in Table \ref{tab:k*pirhopi}~\cite{WagnerThesis,HFAG}. The CP asymmetries in $B^+\to\rho^+\pi^0$ and $B^+\to\rho^0\pi^+$ are consistent with zero within errors, and will 
be taken to vanish at this point. 
We quote $B^+\to \rho\pi$ amplitudes in units of $10^{-3}$, given by square roots of central values for
branching ratios divided by the lifetime ratio $\tau_B$.
The relative phase between these two amplitudes, which is dominantly a strong phase as shown in 
(\ref{rho+pi0_approx}), will be denoted  by
\beq\label{phi}
\phi \equiv {\rm Arg}[A(\rho^0\pi^+)A^*(\rho^+\pi^0)]~.
\eeq
Omitting an overall phase of $A(\rho^+\pi^0)$ we obtain numerically: 
\beq\label{A3/2_num}
3A_{3/2} = 1.05 - 0.56e^{i(\phi - \gamma)}~,~~~~
3\bar A_{3/2} = 1.05 - 0.56e^{i(\phi + \gamma)}~,
\eeq
where $\bar A_{3/2}$ is the corresponding amplitude for $\bar B^0$ decays.

The phase difference $\phi$ is measurable by constructing geometrically an isospin pentagon 
for the five $B^{0,+}\to 3\pi$ decay amplitudes~\cite{Lipkin:1991st,Gronau:1991dq}. 
The measured CP-averaged $B\to \rho\pi$ branching ratios are consistent 
with an approximately flat pentagon~\cite{Gronau:2010dd} which would correspond to 
$\phi\simeq 0$. However, these branching ratios permit also a non-flat pentagon.
Moreover, large values of $\phi$ cannot be excluded because of sizable
experimental errors~\cite{Zhang:2004wza,Aubert:2007py}.
Theoretically, one expects this phase to be small. QCD factorization predicts its suppression
by $\alpha_s(m_b)$ and $1/m_b$. Taking $\phi=0$ and using a value 
$\gamma=65^\circ$ favored by fits to CKM parameters~\cite{Charles:2004jd,Bona:2009ze},
one obtains
\beq\label{A3/2max}
3|A_{3/2}| = 3|\bar A_{3/2}| = 0.96~,~~~~{\rm for}~\phi=0~.
\eeq
For nonzero values of $\phi$ one of these amplitudes decreases while the other increases.
For instance,
\beq\label{90}
3|A_{3/2}| = 0.59~,~~~3|\bar A_{3/2}| = 1.57~,~~~~{\rm for}~\phi=90^{\circ}~.
\eeq
The maximal value of $3|A_{3/2}|$ (or $3|\bar A_{3/2}|$) is 1.61. 
  
In order to calculate $|\Delta\Phi|$ and $|\Delta\overline{\Phi}|$ the above values of 
$3|A_{3/2}|$ and $3|\bar A_{3/2}|$ may be combined with $|A(K^{*+}\pi^-)|$, 
$\s |A(K^{*0}\pi^0)|$ and their charge-conjugates, also expressed in units of $10^{-3}$.  
Using central values of corresponding 
branching ratios in Table I and neglecting CP asymmetries in these processes, one has 
\beq\label{K*pi}
|A(K^{*-}\pi^+)| = |A(K^{*+}\pi^-)| = 2.86~,~~~~
\s |A(\overline{K}^{*0}\pi^0)| = \s |A(K^{*0}\pi^0)| = 2.57~.
\eeq
Comparing the smaller amplitudes (\ref{A3/2max}) for $\phi=0$ with the larger amplitudes 
(\ref{K*pi}) we conclude there is a strong destructive interference
in (\ref{A3/2}) and in its charge-conjugate, corresponding to phase differences
\beq\label{Phi}
|\Delta\Phi| = |\Delta\overline{\Phi}| = 161^\circ~.
\eeq
For $\phi\ne 0$ one of this phases becomes larger than this value while the other phase
becomes smaller reaching a minimum value of $146^\circ$.
\bigskip

{\bf V. INCLUDING SU(3) BREAKING AND PENGUINS IN $B\to\rho\pi$}

\medskip
The values of $I=3/2$ amplitudes (\ref{A3/2max}), (\ref{90}) and the phases (\ref{Phi})
were obtained neglecting several corrections.  These include
penguin amplitudes which have been neglected in (\ref{rho+pi0}) and
consequently in (\ref{3A3/2approx}), and effects of SU(3) breaking in
relations between tree amplitudes in $\Delta S=1$ and $\Delta S=0$ decays.
Including these corrections, Eq.~(\ref{3A3/2approx}) is now replaced by
\bea\label{3A3/2fin}
3A_{3/2} &=& \tl\s\left( A(\rho^+\pi^0) R_1 - 0.61e^{-i\gamma}A(\rho^0\pi^+) R_2\right) \\
&+& \tl (1 + 0.61 e^{-i\gamma}) (A(\bar K^{*0} K^+) - A(K^{*+} \bar K^0)) ~.\nonumber
\eea
$R_{1,2}$ are SU(3) breaking parameters while the second line describes penguin
contributions.  We do not include similar SU(3) breaking factors in the latter
contributions. We checked that such factors  would have a very small effect on
constraining $\Delta\Phi$ and $\Delta\overline{\Phi}$ once uncertainties in
penguin amplitudes are maximized as discussed below.  Although in the above
derivation we seem to have neglected annihilation amplitudes and penguin
contributions $P_{uc,P}$ and $P_{uc,V}$ involving a CKM factor
$\lambda_u^{(s)}$, Eq.\ (\ref{3A3/2fin}) is exact in the SU(3) limit
$R_1=R_2=1$ and does not neglect any amplitude.

We start by discussing the uncertainty caused by neglecting 
the contribution of penguin amplitudes.
As shown in (\ref{P_P}), $P_P$ contributes 
to $A(B^+\to\rho^+\pi^0)$ about $20\%$ of its magnitude. One may assume 
$|P_V|\simeq |P_P|$~\cite{Chiang:2003pm} on the basis of approximately 
equal branching ratios measured for $B^+\to\ K^0\rho^+$ and $B^+\to 
K^{*0}\pi^+$~\cite{HFAG}. To be most 
conservative we will maximize the uncertainty caused by the combination $P_V-P_P$ by 
assuming that the two penguin amplitudes involve a relative minus sign, 
$P_V \simeq -P_P$~\cite{Lipkin:1990us}. 
Thus, neglecting $P_V - P_P$ in the two $B^+\to \rho\pi$ amplitudes (\ref{rho+pi0}) introduces
a maximal uncertainty of  about $40\%$ in each  amplitude.
Including the CKM factors in (\ref{3A3/2fin}), we find that the penguin amplitudes
may contribute at most 50\% of the contribution of the first line in Eq.~(\ref{3A3/2fin}).

In the presence of these penguin contributions the phase $\phi$ defined in (\ref{phi})  is not 
a purely CP-invariant strong phase as we have assumed when obtaining the structure 
(\ref{A3/2_num}).  Denoting
\beq
\bar\phi \equiv {\rm Arg}[A(\rho^0\pi^-)A^*(\rho^-\pi^0)]~,
\eeq
$\bar\phi$ now replaces $\phi$ in the expression for $3\bar A_{3/2}$. In general
one has $\bar\phi \ne \phi$. The difference between these two phases is suppressed by
the ratio of penguin and tree amplitudes in $B^+\to\rho\pi$. As mentioned, $\phi$ and $\bar\phi$ 
are measurable by constructing  the $B\to\rho\pi$ isospin pentagons for $B$ and $\bar B$.  

SU(3) breaking in $T_P + C_V$ and $T_V + C_P$ may be estimated using naive factorization.
We use this estimate as an example illustrating the small effect of SU(3) breaking on the 
values of $\Delta\Phi$ and $\Delta\overline{\Phi}$. 
In $B\to K^*\pi$ one has,
\bea
T_P + C_V & \propto & a_1f_{K^*}F^{B\pi}_0 + a_2f_\pi A^{BK^*}_0~,\nonumber\\
T_V + C_P & \propto & a_2f_{K^*}F^{B\pi}_0 + a_1f_\pi A^{BK^*}_0~,
\eea
where~\cite{Buchalla:1995vs}
\beq
a_1 = c_1 + c_2/3~,~~~~a_2 = c_2 + c_1/3~,~~~~c_1=1.079~,~~~~c_2=-0.178~,
\eeq
and $f_\pi=131$ MeV, $f_{K^*}=218\pm 4$ MeV, $F^{B\pi}_0 = 0.28\pm 0.05, A^{BK^*}_0
= 0.45 \pm 0.07$~\cite{PDG,Beneke:2003zv}. The corresponding tree amplitudes 
for $B\to\rho\pi$ are
given by similar expressions replacing $K^* \to \rho$. The relevant decay constant and form factor
are $f_\rho = 209\pm 1$ MeV and $A^{B\rho}_0 = 0.37 \pm 0.06$.
Using central values for form factors we obtain
\beq\label{SU3br}
R_1 \equiv \frac{(T_P + C_V)_{K^*\pi}}{(T_P + C_V)_{\rho\pi}} = 1.07~,~~~~~
R_2 \equiv \frac{(T_V + C_P)_{K^*\pi}}{(T_V + C_P)_{\rho\pi}} = 1.19~.
\eeq
These SU(3) breaking factors multiply $A(\rho^+\pi^0)$ and $A(\rho^0\pi^+)$ in eq.~(\ref{3A3/2fin}).
As we will see below, these SU(3) breaking corrections do not affect significantly constraints on the
phases $\Delta\Phi$ and $\Delta\overline{\Phi}$. Therefore we will not include in these constraints 
errors caused by uncertainties in $B$ decay form factors.
%
% This is Figure 2
\begin{figure}[h]
\begin{center}
\includegraphics[width=0.45\textwidth]{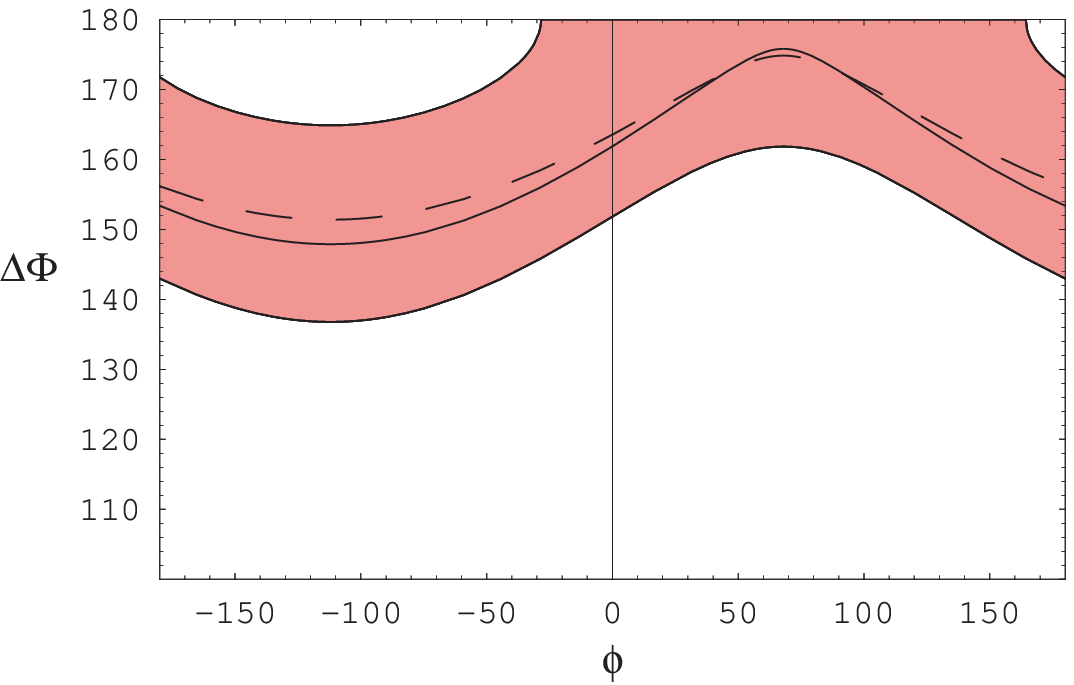}~~~~~
\includegraphics[width=0.45\textwidth]{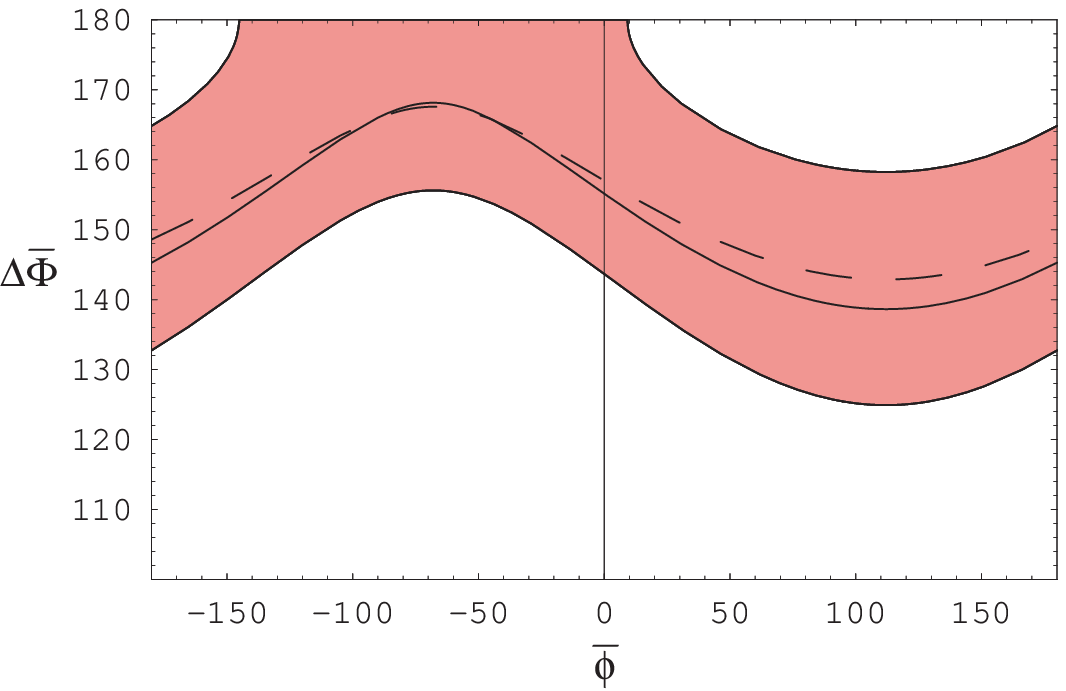}
\end{center}
\caption{Bounds on $\Delta\Phi$ as function of $\phi$ (left), and
on $\Delta\overline{\Phi}$ as function of $\overline{\phi}$ (right). 
Three solid lines describe lower, central and upper values  of 
$\Delta\Phi$ at $1\sigma$. Broken line corresponds to central value for 
the SU(3) symmetric case.
\label{fig:DeltaPhi}}
\end{figure}

We will now study constraints on $|\Delta\Phi|$ and $|\Delta\overline{\Phi}|$ which include 
experimental errors in branching ratios and CP asymmetries in 
$B^+\to\rho^+\pi^0, B^+\to \rho^0\pi^+$, $B^0\to K^{*+}\pi^-, B^0\to K^{*0}\pi^0$.
We take a range for $\gamma$~\cite{Charles:2004jd}, $\gamma = (68 \pm 4)^\circ$,  and theoretical uncertainties from penguin amplitudes in $B^+\to \rho\pi$ decays as described above. All errors are added in quadrature. The numerical SU(3) breaking factors in (\ref{SU3br}) will be used.
Figure \ref{fig:DeltaPhi} shows resulting plots for bounds on $|\Delta\Phi|$ 
and $|\Delta\overline{\Phi}|$ as functions of $\phi$ and $\overline{\phi}$, respectively, in the 
ranges $-180^\circ \le \phi, \overline{\phi} \le 180^\circ$. The three solid lines in each plot describe 
lower, central and upper values at $1\sigma$ for the two phases.
The plots were obtained by taking symmetric errors in $\cos(\Delta\Phi)$ and 
$\cos\Delta\overline{\Phi}$. This assumes that these two variables are linear functions of the 
input parameters.
The broken lines in Figure \ref{fig:DeltaPhi} describe central values 
of $|\Delta\Phi|$ and $|\Delta\overline{\Phi}|$ for the SU(3) 
symmetric case. The few degree difference between the broken line and the 
central solid line  demonstrates the small effect of SU(3) breaking on the allowed ranges of $|\Delta\Phi|$ and $|\Delta\overline{\Phi}|$. 
%
%This is Figure 3
\begin{figure}[h]
\begin{center}
\includegraphics[width=0.45\textwidth]{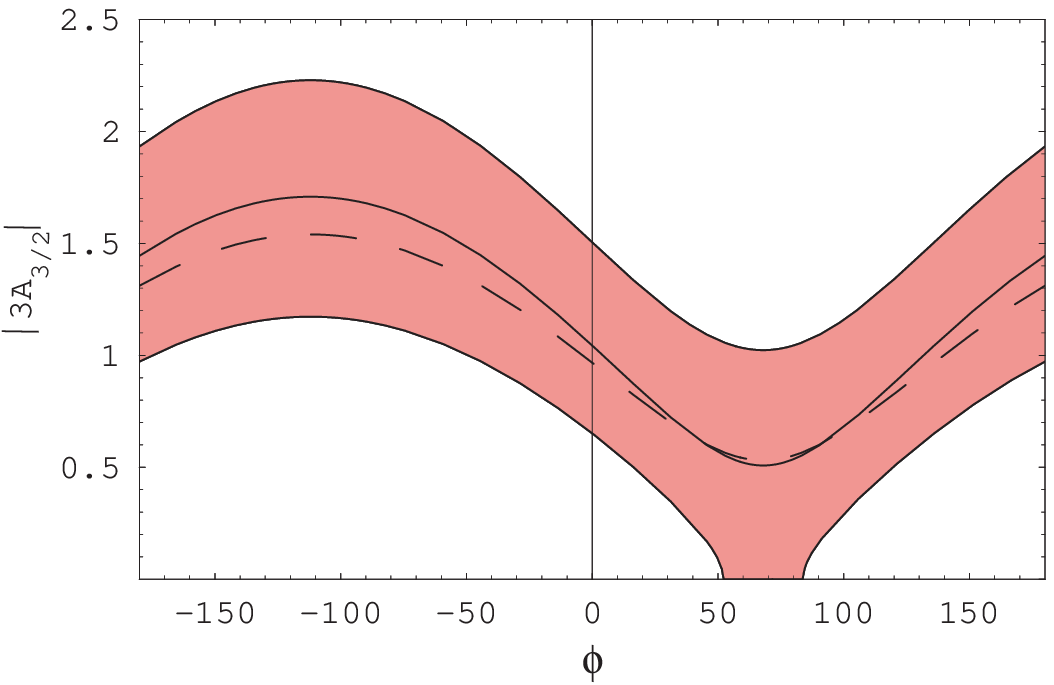}~~~~~
\includegraphics[width=0.45\textwidth]{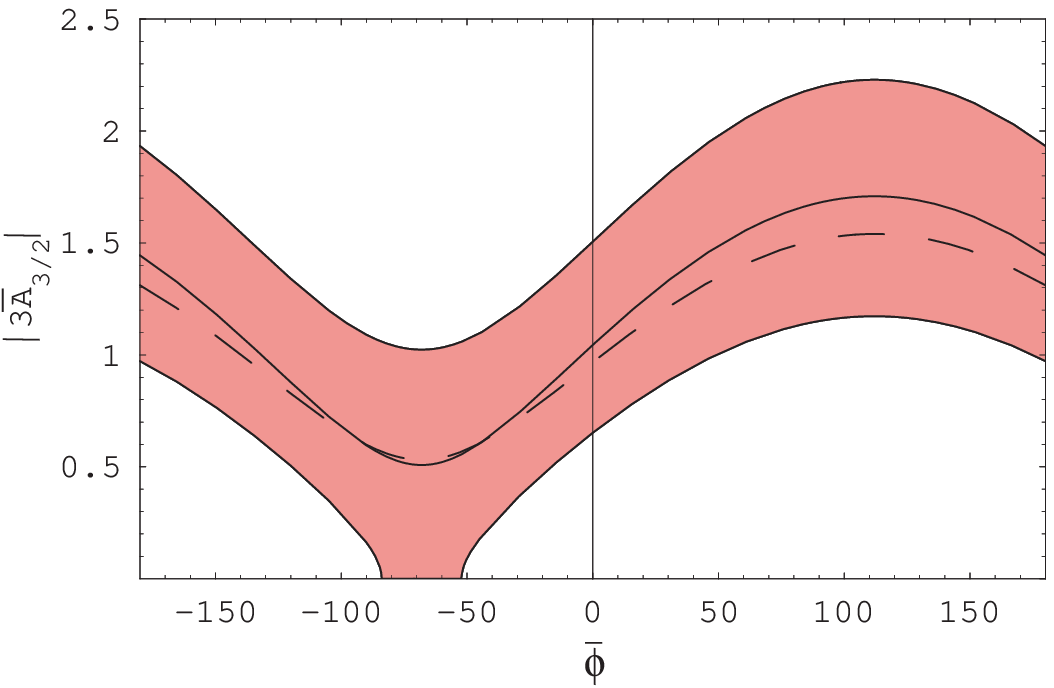}
\end{center}
\caption{Magnitude of isospin amplitude $|3A_{3/2}|$ (left) and $|3\bar A_{3/2}|$
(right) in units of
$10^{-3}$ as function of $\phi$. Lines as in Fig.~\ref{fig:DeltaPhi}.
\label{fig:A3/2}}
\end{figure}

Using Figure \ref{fig:DeltaPhi} and assuming normal distributions 
for $\cos\Phi$ and $\cos\overline{\Phi}$ as functions of the input parameters, we conclude the
following lower limits at $95\%$ confidence level:
\beq\label{SMbounds}
|\Delta\Phi| \ge 131^\circ~,~~~~~~|\Delta\overline{\Phi}| \ge 119^\circ~.
\eeq
These lower bounds correspond to the minimal values of 
$|\Delta\Phi|, |\Delta\overline{\Phi}|$ in 
Figure \ref{fig:DeltaPhi}
which are obtained at $\phi=-180^\circ + \gamma, \bar\phi=180^\circ - \gamma$.
The bounds should be considered conservative as  the magnitudes of the measurable
phases $\phi$ and $\bar\phi$ are not expected to be larger than $90^\circ$.

For completeness, we plot in Figure \ref{fig:A3/2} the
predicted amplitudes 
$3|A_{3/2}|$ and $3|\bar A_{3/2}|$ as functions of $\phi$ and $\bar\phi$, respectively.
Amplitudes in units of $10^{-3}$ are given by square roots of corresponding
branching ratios. 
We note that the two $I=3/2$ amplitudes are different from zero except for restricted
ranges of the phases $\phi$ and $\bar\phi$, $\phi \sim 50^\circ - 80^\circ$, 
$\bar\phi  \sim (-80^\circ) - (-50^\circ)$.
\bigskip

{\bf VI. CONCLUSION: COMPARISON WITH BABAR RESULTS}
%
% This is Table 2
\begin{table}[h]
\caption{Four solutions for $\Delta\Phi' \equiv \Delta\Phi - \pi$ and
$\Delta\overline{\Phi'}\equiv \Delta\overline{\Phi} - \pi$ with minimum values
of the negative likelihood function (NLL) measured in $B\to K^{\pm}\pi^{\mp}
\pi^0$.  Statistical and systematic errors are added in quadrature.  The first
three values in each column are taken from~\cite{Aubert:2008zu}. The last three
values are the results of  a very recent update~\cite{WagnerThesis}.
\label{tab:PhiPhibar}}
\begin{center}
\begin{tabular}{c c c c c c} \hline \hline
~&~ &  Solution I & Solution II & Solution III & Solution IV\\ \hline\hline
Ref.~\cite{Aubert:2008zu} & $\Delta\Phi'$ & $(-21 \pm 35)^\circ$ & $(-134 \pm 30)^\circ$
& $(-22 \pm 30)^\circ$ & $(-139 \pm 30)^\circ$ \\
~ & $\Delta\overline{\Phi'}$ & $(-5 \pm 34)^\circ$ & $(-5 \pm 33)^\circ$ &
$(-163 \pm 33)^\circ$ & $(-163 \pm 33)^\circ$ \\
~ & $\Delta({\rm NLL})$ & $0$ & $3.94$ & $7.77$ & $10.57$ \\ \hline
Ref.~\cite{WagnerThesis} & $\Delta\Phi'$ & $(-22 \pm 39)^\circ$ & $(-139 \pm 40)^\circ$  
& $(-22 \pm 39)^\circ$ & $(-140 \pm 40)^\circ$ \\
~ & $\Delta\overline{\Phi'}$ & $(-5 \pm 36)^\circ$ & $(-4 \pm 36)^\circ$ & 
$(-163 \pm 35)^\circ$ & $(-163 \pm 35)^\circ$ \\
~ & $\Delta({\rm NLL})$ & $0$ & $5.43$ & $7.04$ & $12.33$ \\ \hline
\end{tabular}
\end{center}
\end{table}

We now compare our lower bounds on $|\Delta\Phi|$ and $|\Delta\overline{\Phi}|$
with values reported by Babar in Ref.~\cite{Aubert:2008zu} and in a recent
update~\cite{WagnerThesis}.  Performing a maximum likelihood fit to 4583 $B\to
K^{\pm}\pi^{\mp}\pi^0$ events, four solutions were found for 
$\Delta\Phi' \equiv \Delta\Phi - \pi$ and $\Delta\overline{\Phi'}\equiv \Delta
\overline{\Phi} - \pi$ with minimum values of the negative likelihood function (NLL).
Results of the two analyses are presented in Table \ref{tab:PhiPhibar}, quoting
for each  of the four solutions values for $\Delta\Phi', \Delta\overline{\Phi'}$
and $\Delta({\rm NLL})$, the difference in units of NLL with respect to the
most likely solution (I).  We will compare our bounds to the updated results in
Ref.~\cite{WagnerThesis}.

Solution I with the highest probability favors small values of $\Delta\Phi'$ and 
 $\Delta\overline{\Phi'}$ consistent with zero, or large values of $\Delta\Phi$ and 
 $\Delta\overline{\Phi}$ 
 near $180^\circ$ in agreement with our lower bounds (\ref{SMbounds}). 
 The next likely solution II and III (disfavored by $3.3\sigma$ and $3.8\sigma$) 
 involve one large phase and one small phase, while the most unlikely solution 
 (disfavored by $5\sigma$) consists of large values for both $\Delta\Phi'$ and 
 $\Delta\overline{\Phi'}$.
The highly favored Solution I, using a different convention than ours for the 
two phases, is in agreement with our bounds, corresponding to destructive 
interference between $A(K^{*0}\pi^0)$ and $A(K^{*+}\pi^-)$ in (\ref{A3/2}) and 
between their charge-conjugates.

\begin{figure}
\includegraphics[width=0.48\textwidth]{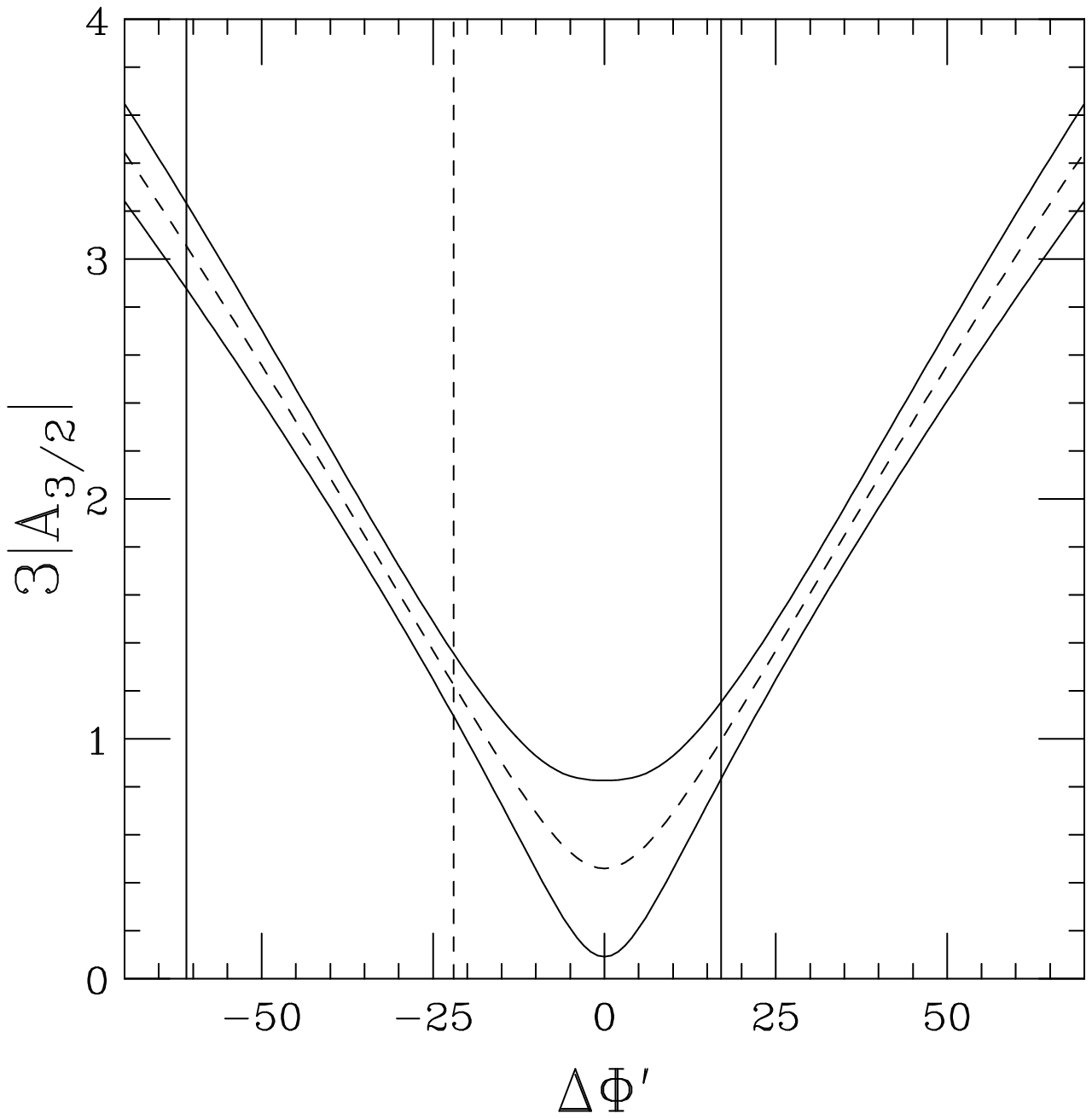}
\includegraphics[width=0.48\textwidth]{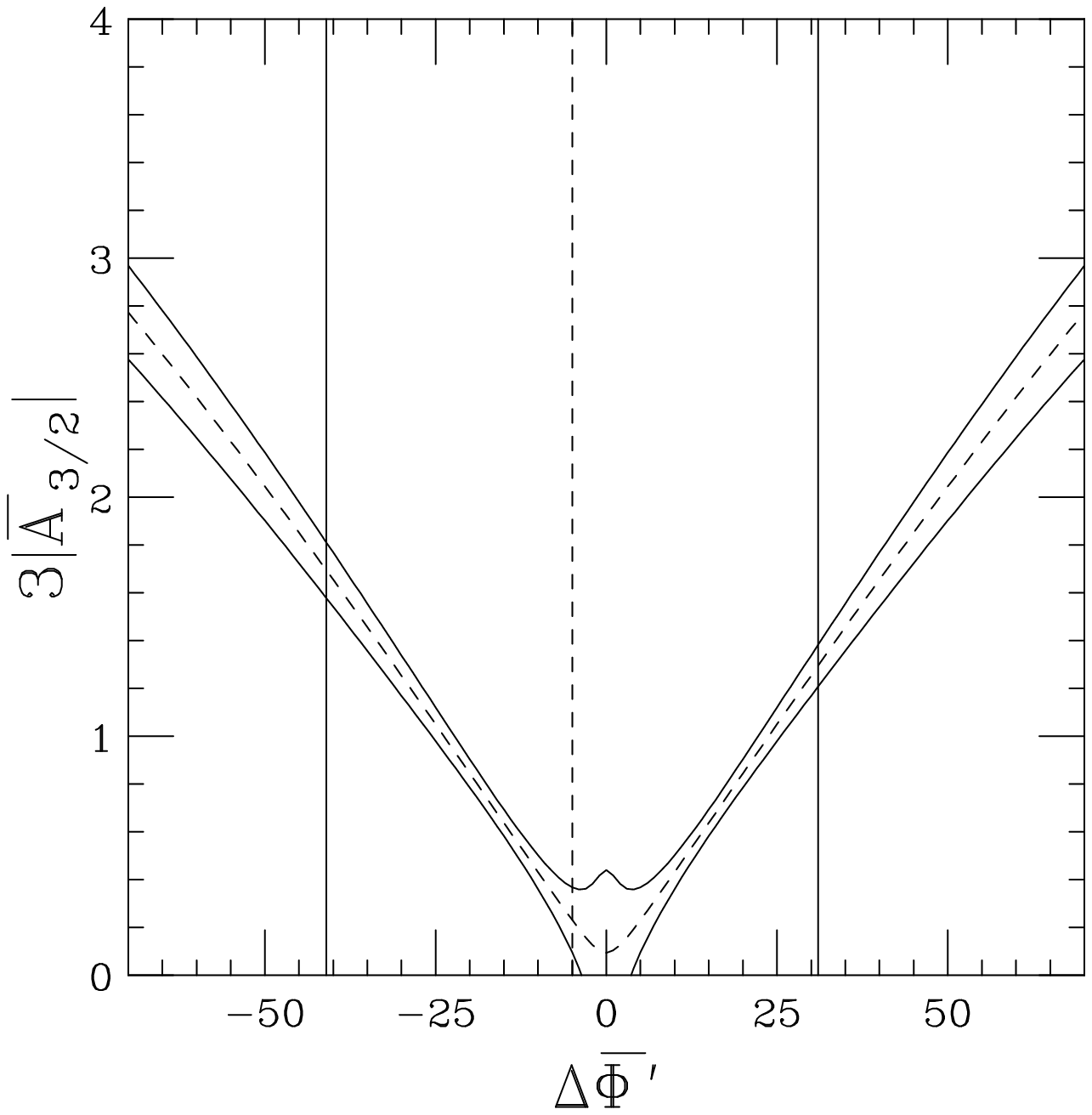}
\caption{Magnitudes of $I=3/2$ amplitudes as functions of relative phases
between $K^{*+} \pi^-$ and $K^{*0} \pi^0$ amplitudes, extracted from
$B\to K^*\pi$  branching ratios and asymmetries given in Table \ref{tab:k*pirhopi}.  
Left:  $3 |A_{3/2}|$; right: $3|\bar A_{3/2}|$.
Vertical lines show central value and $1\sigma$ limits of phases 
$\Delta \Phi'$ or $\Delta \overline{\Phi}'$ quoted in Ref.\ \cite{WagnerThesis}. 
Curves are shown for central values of branching
ratios and CP asymmetries with bands denoting $1 \sigma$ errors added in
quadrature.
\label{fig:amps}}
\end{figure}

Using $B\to K^*\pi$ branching ratios and CP asymmetries quoted in 
Table~\ref{tab:k*pirhopi} (where Babar and Belle results for $B\to K^{*+}\pi^-$ 
have been averaged), values of $\Delta\Phi'$ and 
$\Delta\overline{\Phi'}$ for the favored Solution I~\cite{WagnerThesis}, 
and assuming no correlations between these measurements, we calculate for
central values of branching ratios and CP asymmetries
\beq
3|A_{3/2}| = 1.22^{+1.83}_{-1.22}~,~~3|\bar A_{3/2}| = 0.23^{+1.46}_{-0.23}~,
\eeq
where the errors are due to the uncertainties in $\Delta \Phi'$ and $\Delta
\overline{\Phi}'$.  The dependence of these amplitudes on $\Delta \Phi'$ or
$\Delta \overline{\Phi}'$ and on errors in branching ratios and CP asymmetries
is illustrated in Fig.\ \ref{fig:amps}.

% This is Figure 4
The values of the two isospin 3/2 amplitudes are consistent with zero within large
errors.  Improvement in errors on the relative phases $\Delta \Phi'$ and
$\Delta \overline{\Phi}'$ (depending on their values) may be able to permit
determination of $3|A_{3/2}|$ and $3|\bar A_{3/2}|$ with sufficient accuracy
to constrain their relative phase so as to provide a new constraint on CKM 
parameters~\cite{Ciuchini:2006kv,Gronau:2006qn}.
%MG add 2 sentences
Also, the  CP rate asymmetry, $\Delta((K^*\pi)_{I=3/2}) \equiv (3|\bar A_{3/2}|)^2 - 
(3|A_{3/2}|)^2$, has been shown to be equal to a sum combining eight CP rate 
asymmetries in all possible $B\to K^*\pi$ and $B\to \rho K$ decays~\cite{Gronau:2010dd}.
A potential violation of this sum rule would provide evidence for New Physics.

%MG add sentence
Improvements in the measurements of $|A_{3/2}|, \Delta\Phi$ and their charge-conjugates  
may be achieved in the near future.
The latest results for $B\to K^\pm\pi^\mp\pi^0$ published by the Belle collaboration 
used a data sample from an integrated luminosity of only 
$78$ fb$^{-1}$~\cite{Chang:2004um}. By now 
Belle has accumulated about ten times more data for this decay mode,
approximately twice the amount studied by Babar.  Belle should be encouraged 
to analyze their full set of data in order to improve the measurements of 
$3|A_{3/2}|, \Delta \Phi$ and their charge-conjugates.
\bigskip

We are grateful to Andrew Wagner and Jure Zupan for useful discussions.
M. G. and J. L. R. wish to thank the Galileo Galilei Institute for Theoretical
Physics for hospitality and the INFN for partial support during the completion
of this work.  This work was supported in part by the United States Department
of Energy under Grant No.\ DE-FG02-90ER40560.

\end{document}